\documentclass[prx,twocolumn,showpacs,superscriptaddress]{revtex4-1}

\usepackage{graphicx}
\usepackage{dcolumn}
\usepackage{bm}
\usepackage{color}

\graphicspath{{fig/}}

\begin{document}

\preprint{APS/123-QED}

\title{
Hidden kagome-lattice picture and origin of high conductivity in delafossite PtCoO$_2$
}

\author{Hidetomo Usui}
\affiliation{Department of Physics, Osaka University, 1-1 Machikaneyama-cho, Toyonaka, Osaka, 560-0043, Japan}
\author{Masayuki Ochi}
\affiliation{Department of Physics, Osaka University, 1-1 Machikaneyama-cho, Toyonaka, Osaka, 560-0043, Japan}
\author{Sota Kitamura}
\affiliation{Max-Planck-Institut f\"ur Physik komplexer Systeme, N\"othnitzer Stra{\ss}e 38, 01187 Dresden, Germany}
\author{Takashi Oka}
\affiliation{Max-Planck-Institut f\"ur Physik komplexer Systeme, N\"othnitzer Stra{\ss}e 38, 01187 Dresden, Germany}
\affiliation{Max-Planck-Institut f\"ur Chemische Physik fester Stoffe, N\"othnitzer Stra{\ss}e 40, 01187 Dresden, Germany}
\author{Daisuke Ogura}
\affiliation{Department of Physics, Osaka University, 1-1 Machikaneyama-cho, Toyonaka, Osaka, 560-0043, Japan}
\author{Helge Rosner}
\affiliation{Max-Planck-Institut f\"ur Chemische Physik fester Stoffe, N\"othnitzer Stra{\ss}e 40, 01187 Dresden, Germany}
\author{Maurits W. Haverkort}
\affiliation{Institute for Theoretical Physics, Heidelberg University, Philosophenweg 19, 69120 Heidelberg, Germany}
\author{Veronika Sunko}
\affiliation{Max-Planck-Institut f\"ur Chemische Physik fester Stoffe, N\"othnitzer Stra{\ss}e 40, 01187 Dresden, Germany}
\affiliation{SUPA, School of Physics and Astronomy, University of St. Andrews, St. Andrews KY16 9SS, United Kingdom}
\author{Philip D. C. King}
\affiliation{SUPA, School of Physics and Astronomy, University of St. Andrews, St. Andrews KY16 9SS, United Kingdom}
\author{Andrew P. Mackenzie}
\affiliation{Max-Planck-Institut f\"ur Chemische Physik fester Stoffe, N\"othnitzer Stra{\ss}e 40, 01187 Dresden, Germany}
\affiliation{SUPA, School of Physics and Astronomy, University of St. Andrews, St. Andrews KY16 9SS, United Kingdom}
\author{Kazuhiko Kuroki}
\affiliation{Department of Physics, Osaka University, 1-1 Machikaneyama-cho, Toyonaka, Osaka, 560-0043, Japan}

\date{\today}

\begin{abstract}
We study the electronic structure of delafossite PtCoO$_2$ to elucidate its extremely small resistivity and high mobility. The band exhibits steep dispersion near the Fermi level despite the fact that it is formed mainly by Pt $d$ orbitals that are typically localized. We propose a picture based on two hidden kagome-lattice-like electronic structure: one originating from Pt $s+p_x/p_y$  orbitals, and the other from Pt $d_{3z^2-r^2}+d_{xy}/d_{x^2-y^2}$ orbitals, each placed on the bonds of the triangular lattice. 
In particular, we find that the underlying Pt $s+p_x/p_y$ bands actually determine the steepness of the original dispersion, so that the large Fermi velocity can be attributed to the large width of the Pt $s+p_x/p_y$ band. 
In addition, the kagome-like electronic structure gives rise to ``orbital-momentum locking" on the Fermi surface, which reduces the electron scattering by impurities. We conclude that the combination of the large Fermi velocity and the orbital-momentum locking is likely to be the origin of the extremely small resistivity in PtCoO$_2$. 
\end{abstract}

\pacs{ }
\maketitle

\section{Introduction}
The past decade has seen considerable attention paid to an unusual series of metals PdCoO$_2$, PdRhO$_2$, PdCrO$_2$ and PtCoO$_2$.\cite{MackenzieReview}
Their strongly two-dimensional conduction takes place in triangular lattice layers of Pd or Pt, separated by layers of edge-sharing transition metal-oxygen octahedra in a three formula unit stacking sequence known as the delafossite structure.
They are particularly notable for their high electrical conductivity, which similar to that of elemental Cu or Ag at room temperature even though their volume carrier density is a factor of three lower.
At low temperatures their mean free paths are as high as tens of microns, opening the way to the investigation of new regimes of electrical transport \cite{Takatsu, Daou, Kikugawa, Moll}.
Recent work on bulk single crystals with well-defined electrical contact geometries defined using focused ion beam sculpting has established the lowest room temperature resistivity among the series to be that of PtCoO$_2$: 1.8 $\mu\Omega$cm \cite{Nandi}.
Intuitively, it is difficult to imagine a three-component oxide having a resistivity this low, and there is a strong motivation to try and understand why this happens.  That is the purpose of this paper.
From the electronic structure point of view, both first principles band structure calculations \cite{Eyert,Noh,Ong} and experiments such as the de Haas--van Alphen \cite{Hicks} and angle resolved photoemission \cite{Noh,Kushwaha}  measurements show a very dispersive band crossing the Fermi level. Although this is consistent with the high conductivity, it is itself puzzling since the orbital projection within first principles calculations show that the band crossing the Fermi level mainly originates from Pt $d$ orbitals (Refs. \cite{Eyert,Ong}, see also Fig.\ref{fig2}), which usually give a narrow band width and a small Fermi velocity. The possibility of a contribution of $s$ orbitals has been discussed in this context \cite{Hicks,MackenzieReview}. 
\begin{figure}
	\includegraphics[width=8cm]{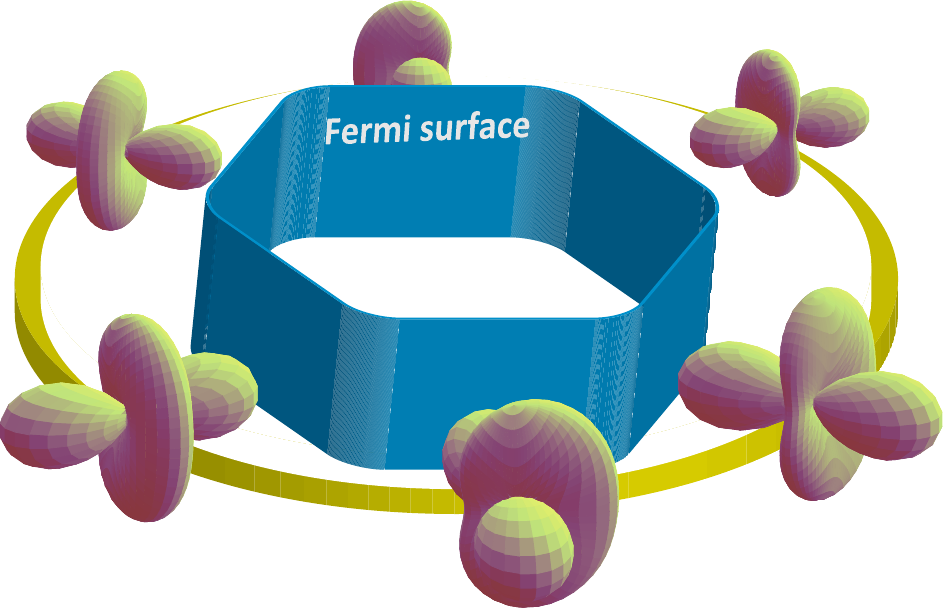}
	\caption{Schematic figure of the orbital-momentum locking. The figure shows how the orbital character varies along the Fermi Surface.}
	\label{fig0}
\end{figure}   

In the present work, we study the electronic structure of PtCoO$_2$, and show that the steep band intersecting the Fermi level is composed of a mixture of 
two hidden kagome-like electronic structures, one originating from Pt $6s$, $6p_x$ and $6p_y$ orbitals, and the other from Pt $5d_{3z^2-r^2}$, $5d_{x^2-y^2}$ 
and $5d_{xy}$. The linear combination of the orbitals forms a basis for a hypothetical atomic orbital on a kagome lattice placed at the bond center of 
the delafossite triangular lattice. In particular, the $s+p_x/p_y$ kagome-like electronic structure has a very large band width of 30 eV, which gives rise to 
the steep dispersion of the band intersecting the Fermi level in the original band structure, despite the fact that $d$-orbital character is much stronger than 
$s+p_x/p_y$ character in this band. Speaking of a kagome lattice (to be precise, a tight-binding model on a kagome lattice with one orbital per site), one may think of 
Dirac cones at the K and K$'$ points, or the presence of a flat band \cite{kagome-band}. Here, however, we will focus on another aspect of the kagome-like electronic 
structure, namely the presence of the quadratic band crossing point~\cite{Sun2009} at the $\Gamma$ point: the kagome lattice is known to have this degeneracy with 
the Berry phase $2\pi$, as a touching of the flat band and one of the dispersive bands. This degenerate point is robust under a six-fold rotational symmetry along 
with an anti-unitary symmetry, so that the present system is expected to have peculiar properties derived from this touching, although the degeneracy is lifted in the actual band structure due to spin-orbit coupling. 
In fact, in the electronic structure of PtCoO$_2$, the orbital character varies along the Fermi surface, as sketched in Fig.\ref{fig0}, giving rise to 
``orbital-momentum locking'', which reduces the rate of the electron scattering by impurities \cite{phonon}. We conclude that the combination of the large Fermi velocity 
and the orbital-momentum locking is likely to be the origin of the extremely small resistivity in PtCoO$_2$. 
\ \\

\section{Band structure}
\begin{figure}
	\includegraphics[width=5cm]{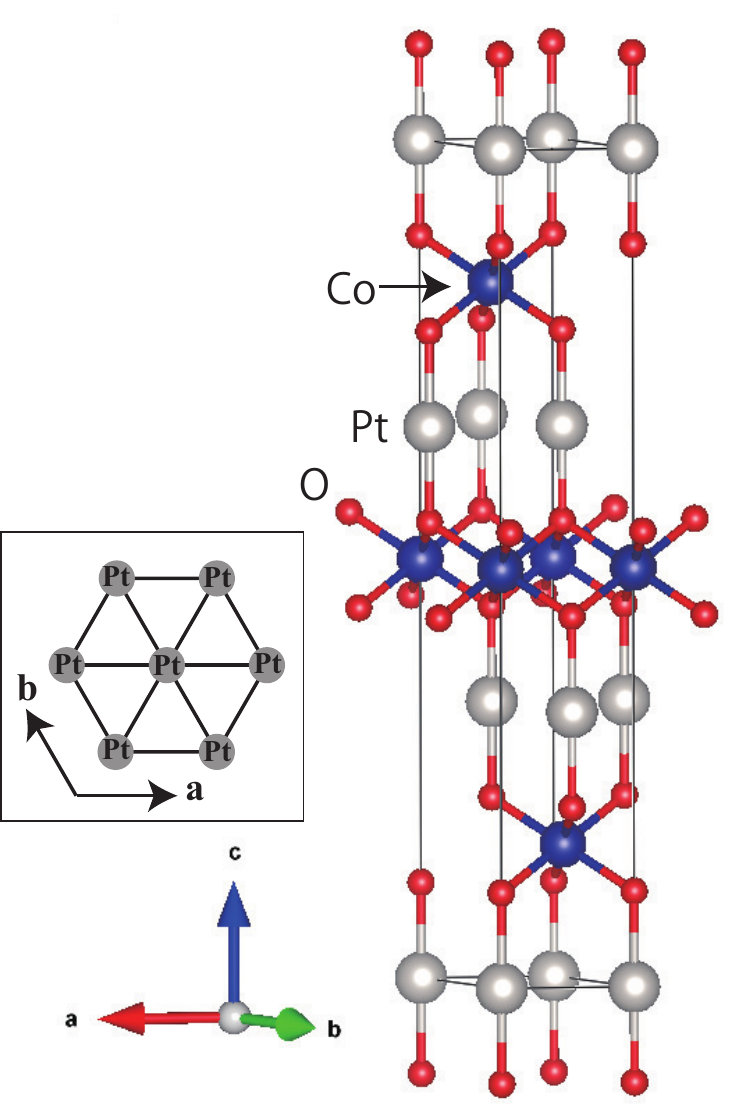}
	\caption{The crystal structure of PtCoO$_2$ depicted using VESTA \cite{VESTA}. The inset shows the triangular lattice of Pt atoms.}
	\label{fig1}
\end{figure}

\begin{figure}
	\includegraphics[width=8cm]{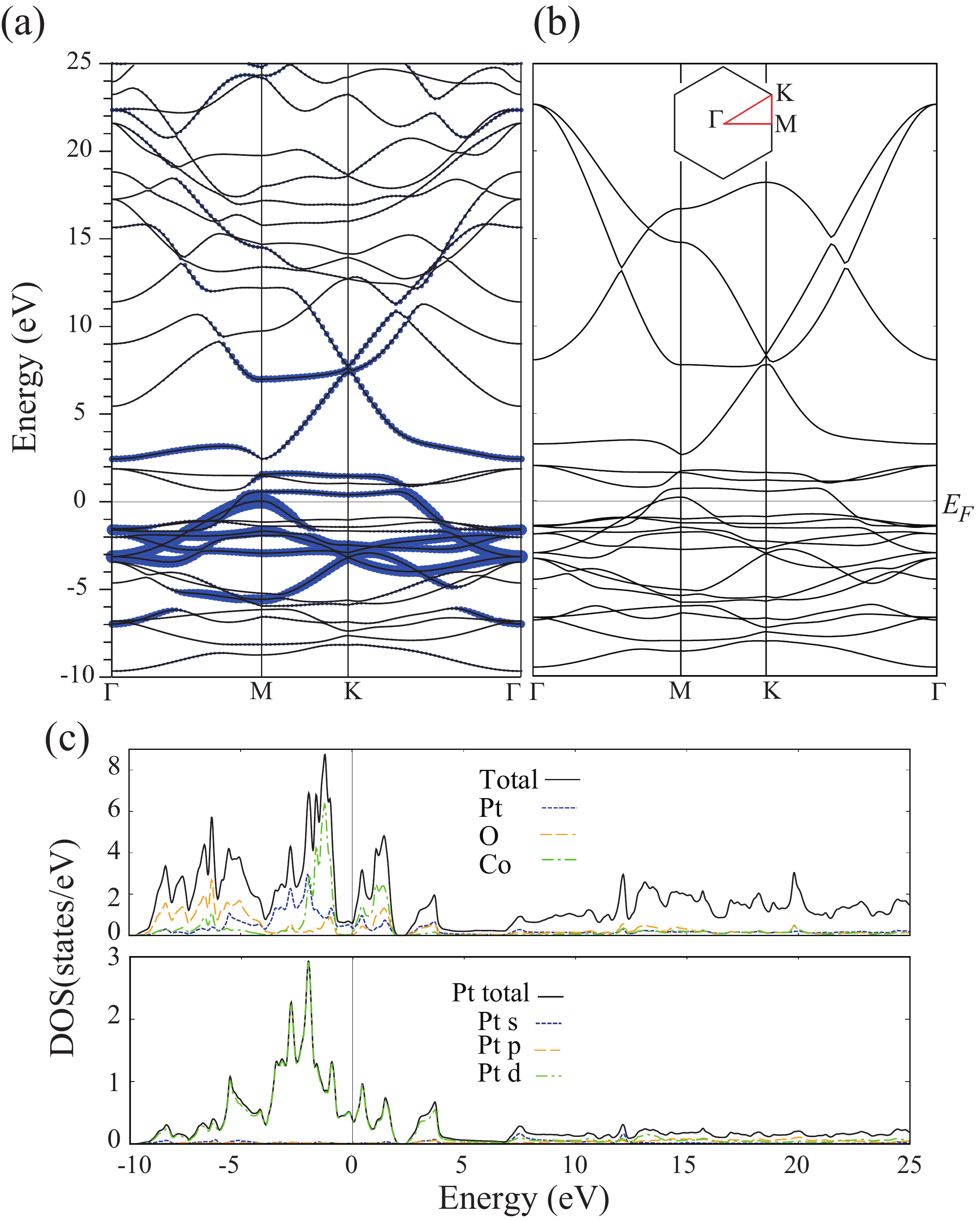}
	\caption{(a) first principles band structure of PtCoO$_2$. The thickness represents the strength of the $d$ orbital character. (b) the band structure of the 20 orbital tight-binding model. (c) the density of states of PtCoO$_2$.}
	\label{fig2}
\end{figure}

The first principles band calculation of PtCoO$_2$, whose crystal structure is shown in Fig. \ref{fig1}, was performed using the WIEN2K package \cite{Wien2k} with the PBE-GGA exchange-correlation functional\cite{PBE} and adopting the lattice parameters obtained in Ref. \cite{Eyert}. The value of $RK_{\rm max}$ is set to 8, and 1,000 k-points are taken for the self-consistent calculation. In our first principles calculation, we have omitted spin-orbit coupling for the sake of the clarity of the argument regarding the hidden kagome-like electronic structure, but we will consider spin-orbit coupling in section \ref{impurity}.
The calculation result is shown in Fig. \ref{fig2}(a) and (c).
This calculation shows that the bands around the Fermi level have strong 
Pt $d$ orbital character, which is expected to give a narrow band width.
However, the band dispersion around the Fermi level is very steep, consistent with that observed experimentally \cite{Kushwaha},
and with its very high room-temperature conductivity..

\section{Orbital decomposition}
\begin{figure}
	\includegraphics[width=8cm]{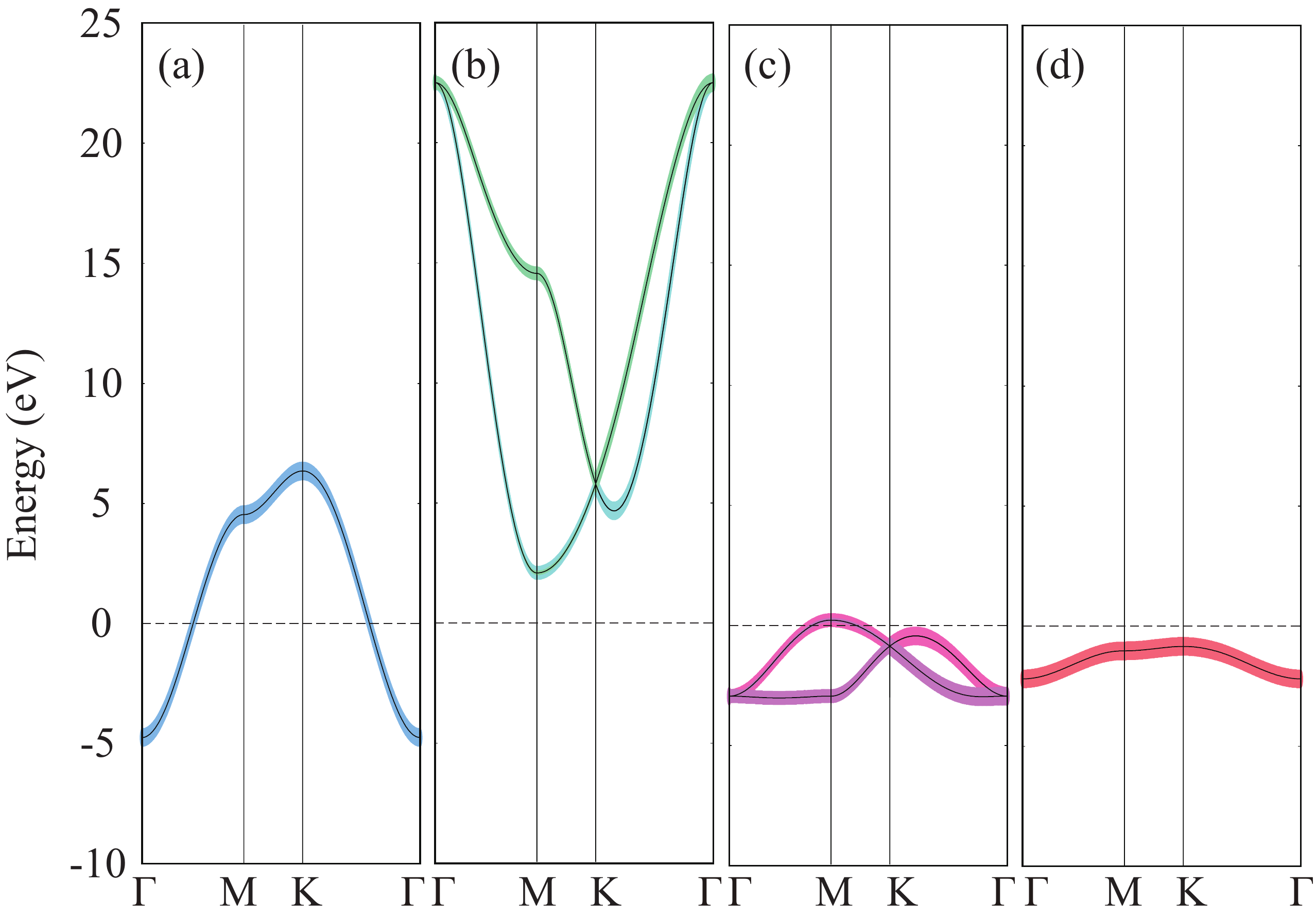}
	\caption{Bands which consist solely of (a) Pt $s$,  (b) Pt $p_x/p_y$, (c) Pt $d_{xy}/d_{x^2-y^2}$, or (d) Pt $d_{3z^2-r^2}$ orbital components, extracted from the 20 orbital model derived from the first principles band structure.}
	\label{fig3}
\end{figure}

\begin{figure}
	\includegraphics[width=8cm]{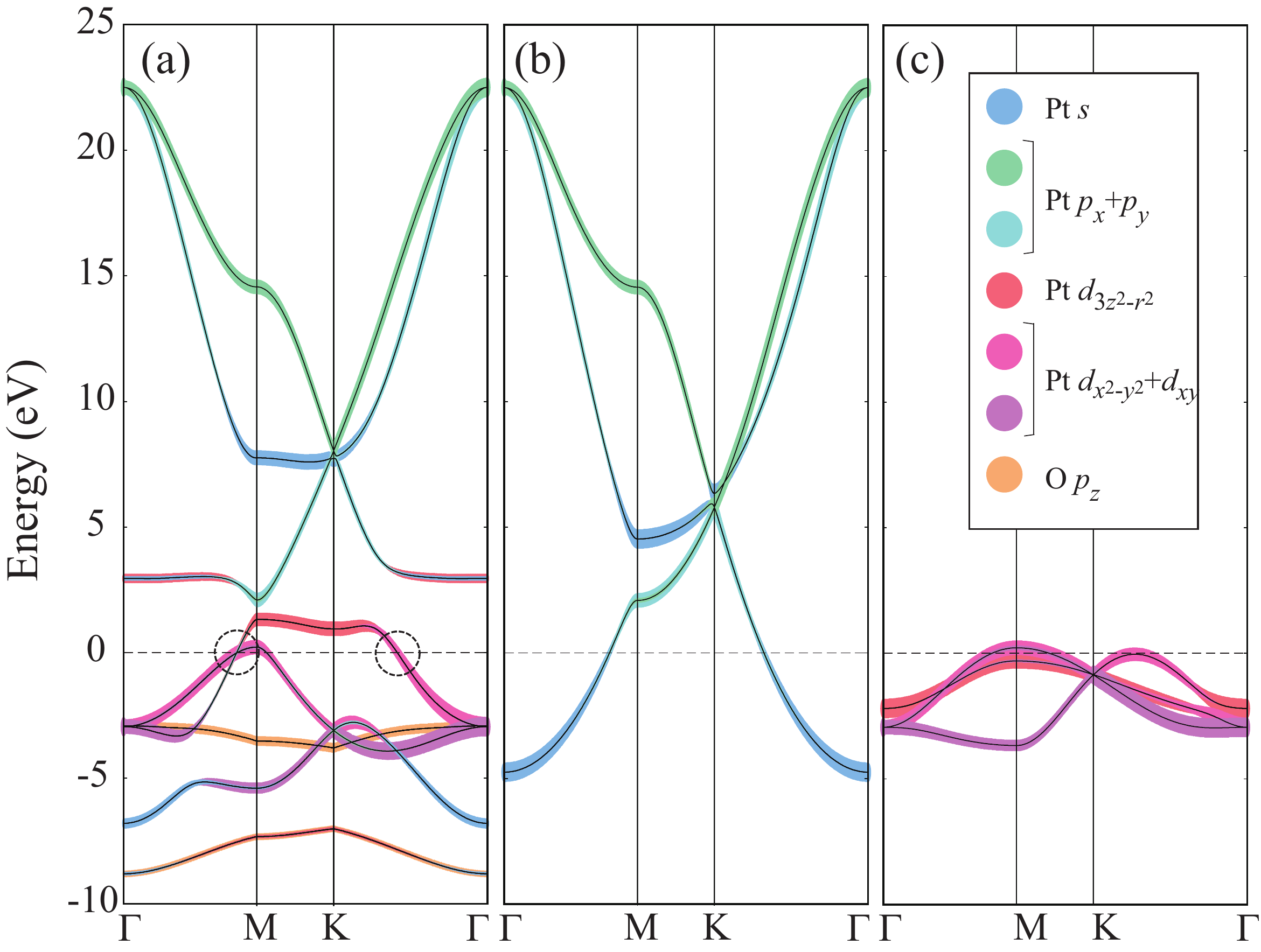}
	\caption{Band structure of models that consist of (a) Pt $s$, $p_x/p_y$, $d_{3z^2-r^2}$, $d_{xy}/d_{x^2-y^2}$, and O$p_z$, (b) Pt $s$ and $p_x/p_y$, (c) Pt $d_{3z^2-r^2}$ and $d_{xy}/d_{x^2-y^2}$ orbitals. Dashed circles in (a) denote the steep band intersecting the Fermi level.}
	\label{fig4}
\end{figure}

In order to understand the origin of this steep dispersion, 
we first construct a 20 orbital tight-binding model,
which consists of 
Pt $s\time 1$, $p\times 3$, $d\times 5$,  Co $d\times 5$, and O $p\times 3$ 
orbitals, exploiting maximally localized Wannier functions \cite{Wannier,w2w}. 
Some of the nearest neighbor hopping integrals obtained are given in table \ref{table1}.
As shown in Fig. \ref{fig2}(b), the tight-binding model accurately reproduces the 
original band structure. From this model, we can extract a hypothetical band structure in which only the hoppings among, say, the Pt $s$ orbital are considered, with no other orbitals mixed. Similar hypothetical band structures can also be obtained for Pt $p_x/p_y$,  Pt $d_{3z^2-r^2}$, or Pt $d_{xy}/d_{x^2-y^2}$ orbitals, as shown in Fig. \ref{fig3}. 
As expected, the $d$ orbital-originated  bands have a narrow band width.
In fact, the mixture of Pt $s$ and $p_x/p_y$ gives rise to a band structure that has very large band width with a steep dispersion, as shown in Fig. \ref{fig4}(b). The steep band intersecting the Fermi level in the original band structure can basically be decomposed into Pt $s+p_x/p_y$ and Pt $d_{3z^2-r^2}+d_{xy}/d_{x^2-y^2}$ orbital components, where the steepness comes from the former, despite the fact that a strong contribution near the Fermi level comes from the latter. 

\begin{table}[htb]
\caption{The value of the nearest neighbor hopping integrals between $s$, $p_x$, $p_y$ and $d_{3z^2-r^2}$ orbitals and the on-site energy of these orbitals.
	$t_1$ and $t_2$ are the nearest neighbor hoppings from (0,0) to (1,0) and (0,1), respectively.
	$(n,m)$ stands for $n{\bf a}$+$m{\bf b}$, where ${\bf a}$ and ${\bf b}$ are the primitive translation vectors shown in Fig. \ref{fig1}. 
	The ``on-site" in the bottom row is the hopping within the same site.\label{table1}}
  \begin{tabular}{|c|c|c|c|c|c|c|} \hline
               &         &  $s$ & $p_x$  & $p_y$ & $d_{3z^2-r^2}$ \\ \hline
               & on-site & 2.41 & 11.31 & 11.31 & -1.30 \\ \hline
 $s$           & $t_1$   &-1.20 & -     & -     & - \\
               & $t_2$   &-1.20 & -     & -     & - \\ \hline
$p_x$          & $t_1$   & 1.87 &  3.25 & -     & - \\
               & $t_2$   &-0.93 &  1.00 & -     & - \\ \hline
$p_y$          & $t_1$   & -    & -     &  0.24 & - \\
               & $t_2$   & 1.62 & -1.30 &  2.50 & - \\ \hline
$d_{3z^2-r^2}$ & $t_1$   &-0.37 & -0.63 & -     &-0.15 \\
               & $t_2$   &-0.37 &  0.32 & -0.55 &-0.15 \\
               & on-site &-0.87 & -     & -     & - \\ \hline
  \end{tabular}
\end{table}

\begin{figure}
	\includegraphics[width=8cm]{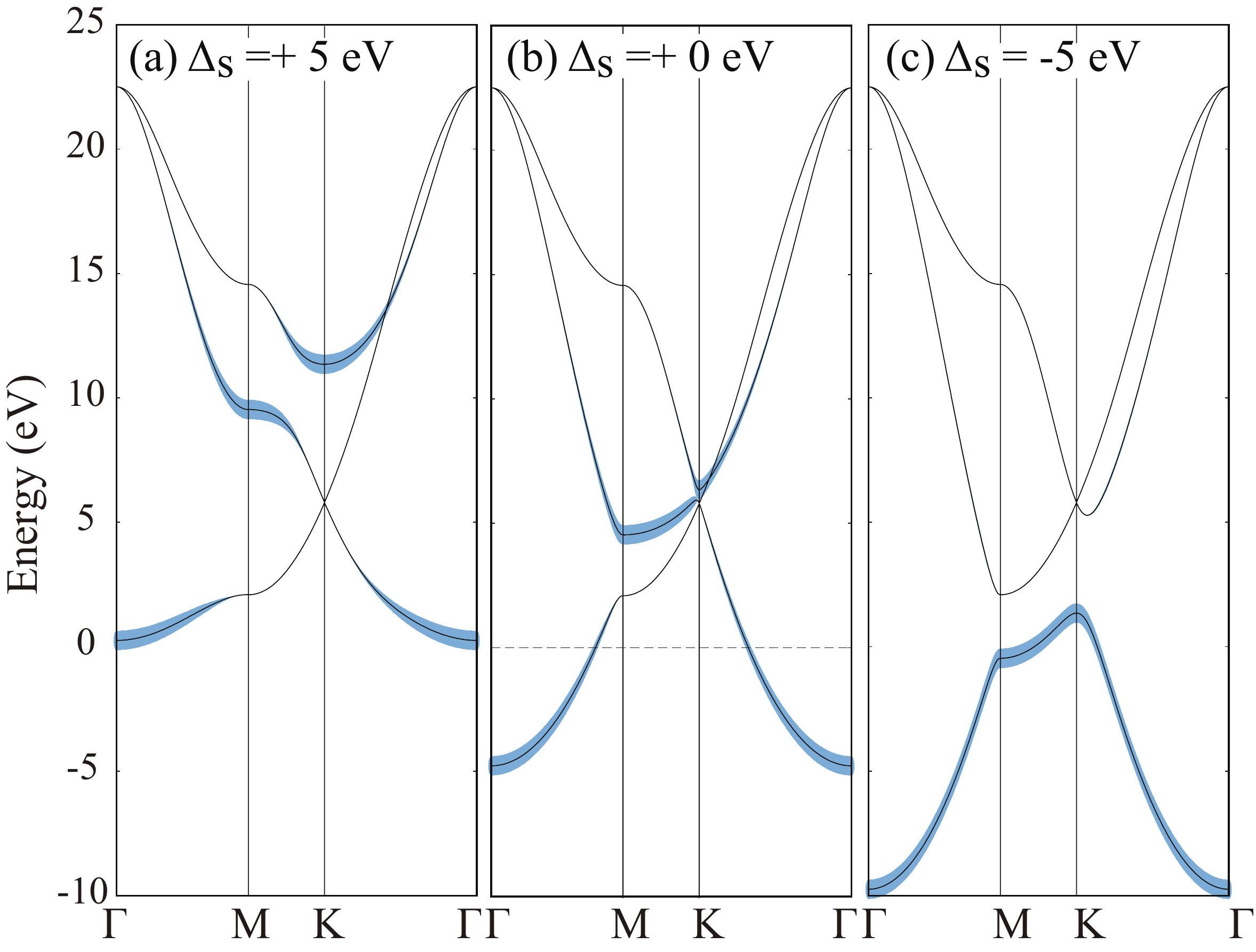}
	\caption{(a)-(c) Evoloution of the $s+p_x/p_y$ band structure upon varying the on-site energy level of the $s$ orbital. The thickness repsents the strength of the $s$ orbital component. 
}
	\label{fig5}
\end{figure}

If we look more closely into the Pt $s+p_x/p_y$ band, it has a Dirac-cone-like feature similar to that of the honeycomb lattice. Figure \ref{fig5} shows how the mixture of $s$ and $p_x/p_y$ orbitals results in a Dirac-cone-like feature by hypothetically varying the $s$-orbital on-site energy; namely, when the $s$ energy level is lowered (Fig. \ref{fig5}(c)),  the $s$ and $p_x/p_y$ bands are clearly separated, but the Dirac cone becomes apparent when the $s$ level is raised and the bands are sufficiently mixed (Fig. \ref{fig5}(a)). Speaking of the honeycomb lattice, the present Pt $s+p_x/p_y$ band has a total band width of nearly 30 eV, which is even larger than that of the graphene. Hence the large group velocity of the band intersecting the Fermi level in the original band structure (dashed circles in Fig. \ref{fig4}(a)) can be traced back to the steep dispersion of the Pt $s+p_x/p_y$ bands.  

\section{Hidden kagome lattice and Orbital Momentum Locking}
\begin{figure}
	\includegraphics[width=8cm]{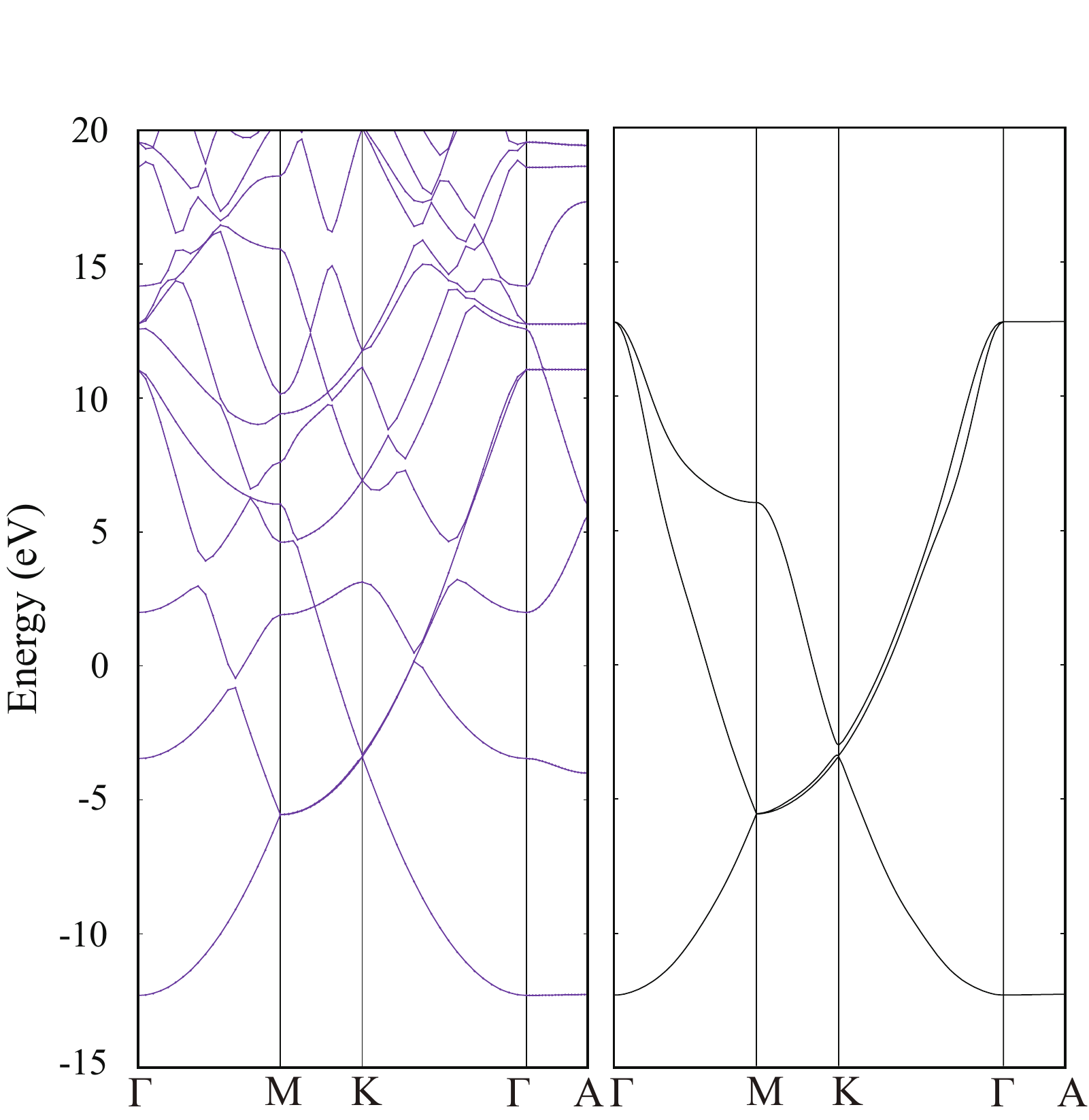}
	\caption{(a) First principles band structure of triangular lattice Si. (b) $s+p_x/p_y$ three orbital model derived from the first principles band structure.}
	\label{fig6}
\end{figure}

\begin{figure}
	\includegraphics[width=7cm]{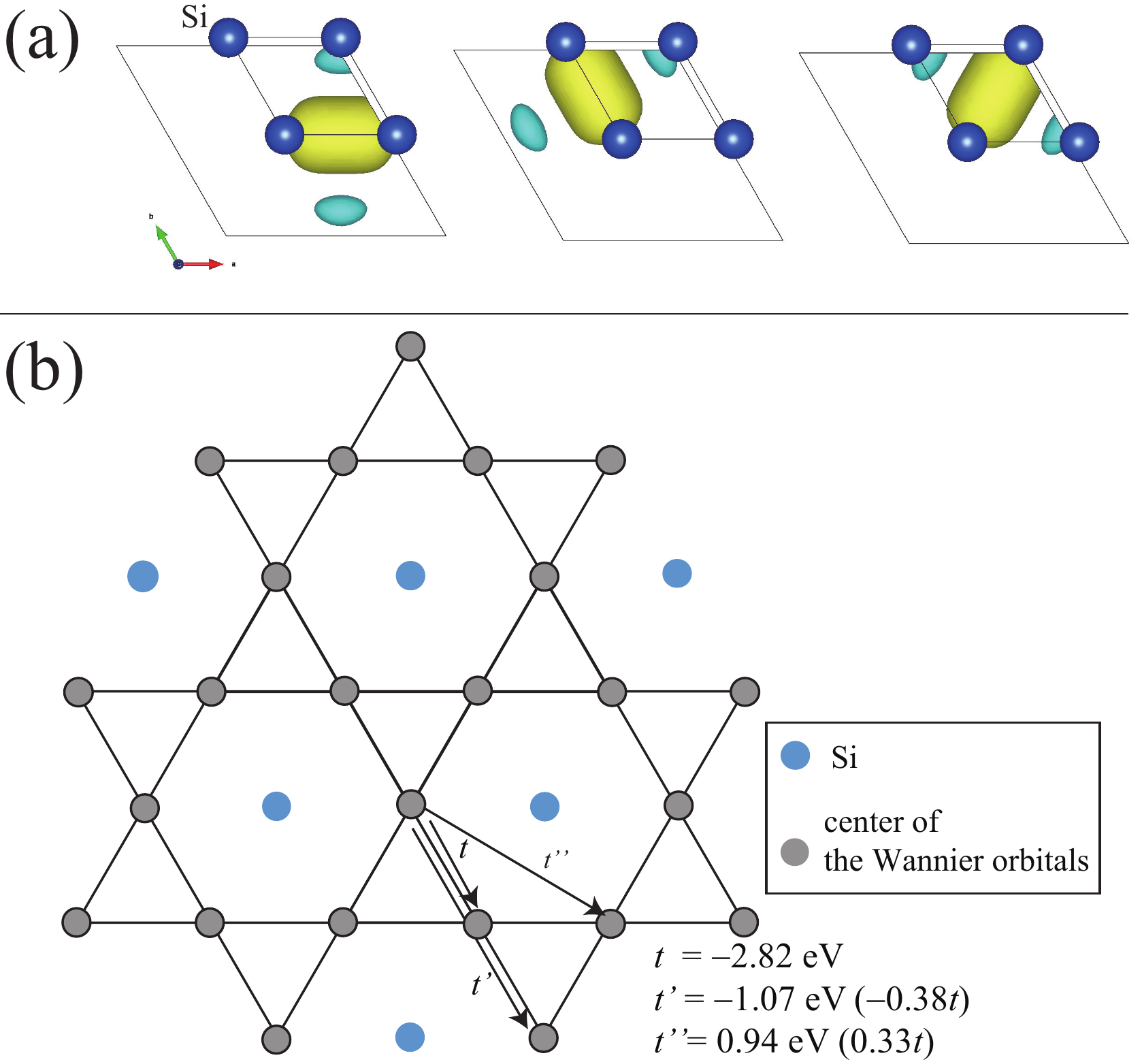}
	\caption{(a) Wannier orbitals of the $s+p_x/p_y$ three orbital model of triangular lattice Si depicted using VESTA \cite{VESTA}. (b) The center of the Wannier orbitals form a kagome lattice.}
	\label{fig7}
\end{figure}

\begin{figure}
	\includegraphics[width=8cm]{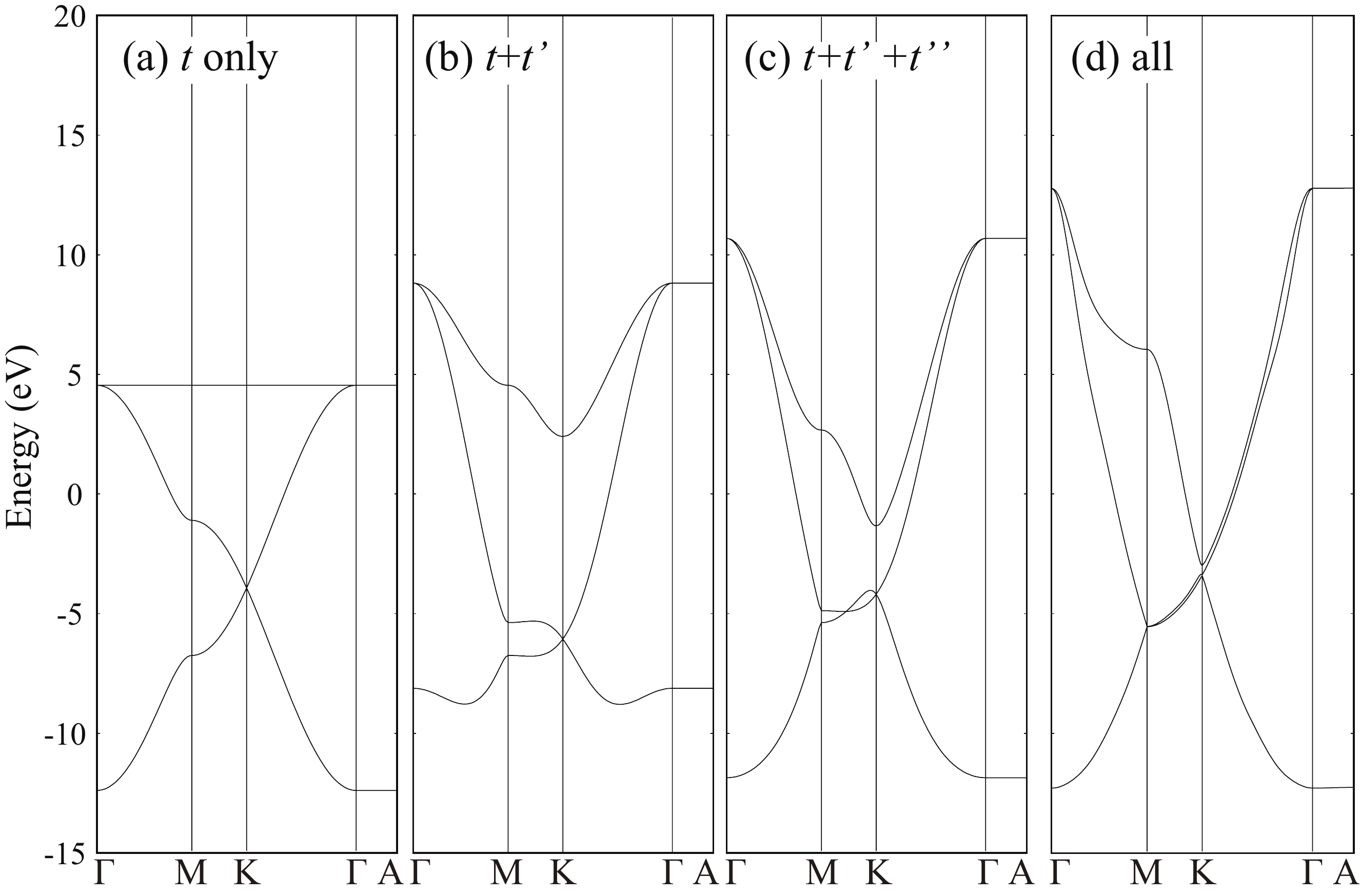}
	\caption{The relation between the kagome lattice with nearest neighbor hopping only and the Si $s+p_x/p_y$ model is shown by adding the distant hoppings one by one. The values of the hopping integrals are shown in Fig. 8.}
	\label{fig8}
\end{figure}

To further understand the origin of this peculiar band structure, we now try to construct a tighbinding model on a triangular lattice for a simpler system that consists of only $s$ and $p_x/p_y$ orbitals with no other bands mixing. To obtain such a model, here we consider a simple hypothetical material, Si on a triangular lattice ($a = 5$ Bohr and $c = 10$ Bohr), where $s$ and $p_x/p_y$ hybridize with no other bands mixing. 
In Fig. \ref{fig6}, we show the original band structure along with the $s+p_x/p_y$ three orbital tight-binding model constructed from maximally localized Wannier functions obtained using VASP \cite{VASP1,VASP2,VASP3} and wannier90 \cite{Wannier} packages. Interestingly, the band structure of this model (Fig. \ref{fig6}(b)) looks very similar to the Pt $s+p_x/p_y$ bands of PtCoO$_2$ (Fig. \ref{fig4}(b)); in the latter the degeneracy of the lower two bands at the M point is lifted, but otherwise the two band structures look just alike. In Fig. \ref{fig7}, we show the Wannier orbitals of the triangular lattice Si, which is constructed with the projection of the $s$ orbital centered at the bond center of two neighboring Si atoms. Thus, the Wannier centers actually form a kagome lattice as shown in Fig. \ref{fig7} \cite{commentLG}. The relation between the present three band model and the tight-binding model on a kagome lattice with nearest neighbor hopping only can be seen by adding distant hoppings one by one as in Fig. \ref{fig8}. When there is only the nearest neighbor hopping, there is a perfectly flat band in addition to the two bands that form Dirac cones, but as the distant hoppings are added, the flat band gets dispersive \cite{commentkagome}, and becomes nearly degenerate with the other two bands at the K point.

\begin{figure}
	\includegraphics[width=6cm]{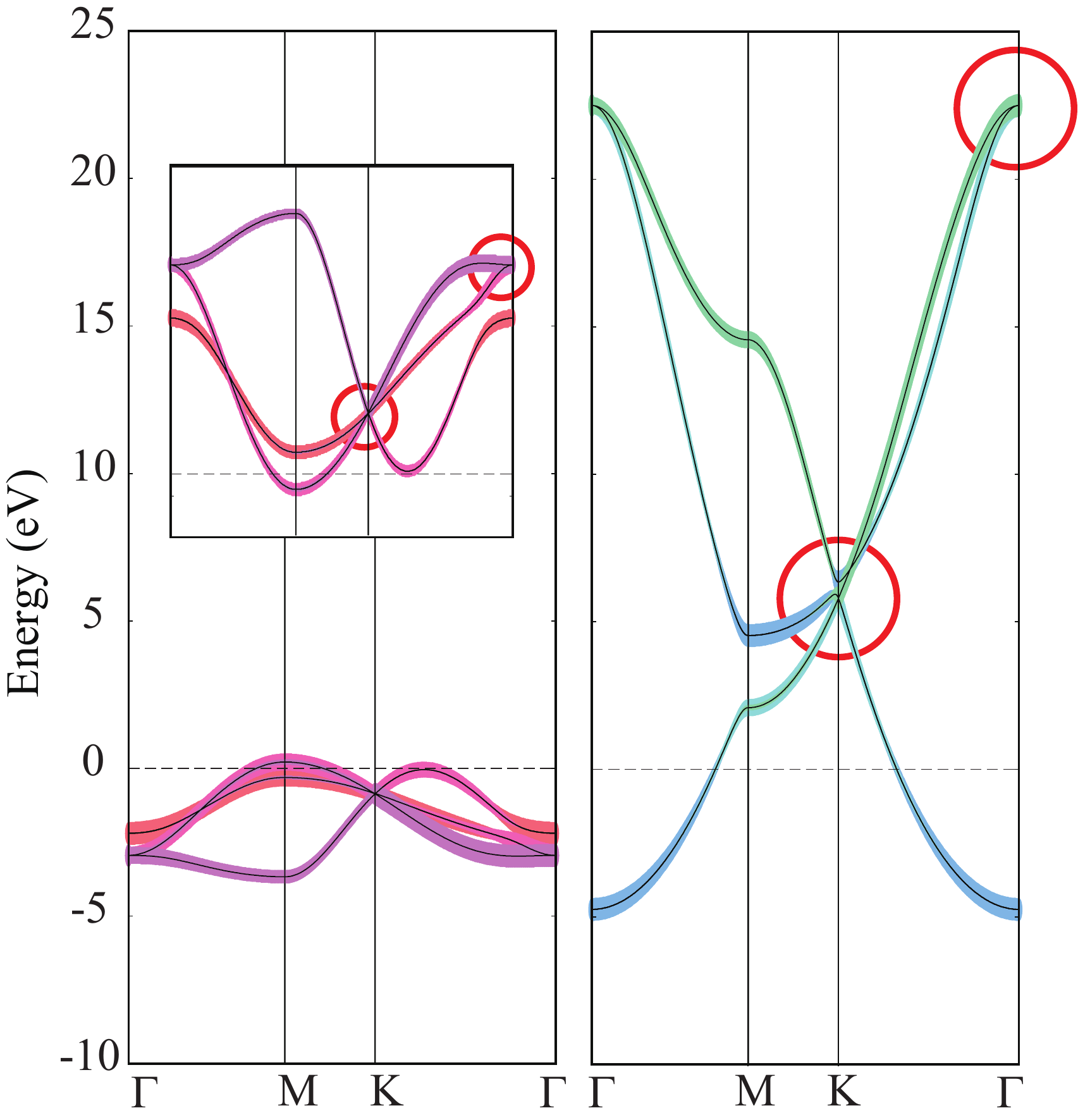}
	\caption{The simlarity between the band structure of Pt $s+p_x/p_y$ (right) and that of Pt $d_{3z^2-r^2}+d_{xy}/d_{x^2-y^2}$ (left) is shown.}
	\label{fig9}
\end{figure}

From the above, we have understood that the steep dispersion of the band intersecting the Fermi level in PtCoO$_2$ originates from the underlying kagome-like electronic structure consisting of Pt $s+p_x/p_y$ orbitals. However, this is not the whole story. If we look closely at the Pt $d_{3z^2-r^2}+d_{xy}/d_{x^2-y^2}$ portion of the band (Fig. \ref{fig4}), which makes a large contribution to the Fermi surface, this itself also has a (strongly deformed) kagome-like electronic structure as seen from the comparison of the reversed blow-up of $d_{3z^2-r^2}+d_{xy}/d_{x^2-y^2}$ and $s+p_x/p_y$ bands in Fig. \ref{fig9}. Namely, there is a Dirac-cone like feature at the K point, and a two-fold degeneracy at the $\Gamma$ point. This may be naturally understood because the $d_{3z^2-r^2}$ ($a_{1g}$ symmetry) and the degenerate $d_{xy}/d_{x^2-y^2}$ ($e_{g}$ symmetry) orbitals are likely to play the same role as $s$ and the degenerate $p_x/p_y$ orbitals, respectively.  

In fact, a relation between the kagome lattice and $3d$ band manifold in a cobaltate Na$_x$CoO$_2$, where Co atoms form a triangular lattice, has also been pointed out in refs. \cite{Koshibae,Alloul1,Alloul2,Lysogorskiy,AlloulRes,sd2Graphene}. In Na$_x$CoO$_2$ also, the $a_{1g}$ and the doubly degenerate $e_g'$ orbitals are the foundation of the relevant band structure. On the other hand, in Na$_x$CoO$_2$, the $s$ and $p$ orbitals do not play any role in the bands near the Fermi level. The difference between Na$_x$CoO$_2$ and PtCoO$_2$ lies in that in the former, the bands around the Fermi level mainly originate from the $3d$ orbitals of the Co atoms, which are surrounded by oxygen atoms that strongly push up the energy level of the widely spread Co $4s$ and $4p$ orbitals.  By contrast, in PtCoO$_2$, the main player is Pt, which by itself forms a layer of triangular lattice.  

\begin{figure}
	\includegraphics[width=6cm]{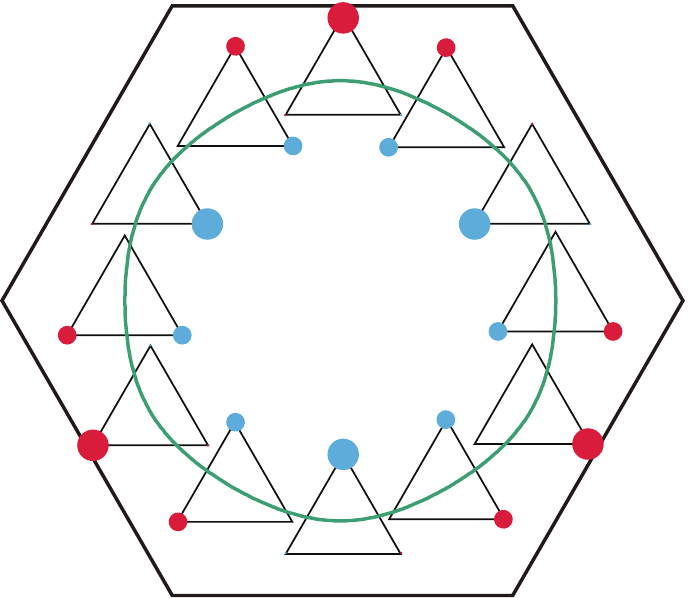}
	\caption{Wavefunction of the original kagome lattice plotted on a real space unit cell. The large circle is a certain energy contour that encircles the $\Gamma$ point, and the wavefunction is plotted along this contour. The radius of the circle represents the weight on each site, and the color denotes the sign (red: negative, blue: positive).}
	\label{fig9-1}
\end{figure}

 Now, as mentioned in the Introduction, a characteric feature of the kagome-like band structure is the Berry phase of $2\pi$ arising from the touching of the two-bands at the $\Gamma$ point. Let us first see this feature directly in the original kagome lattice. We show in Fig. \ref{fig9-1} how the wavefunction, plotted on a real space unit cell,  varies as we move along a certain energy contour around the $\Gamma$ point. It can be seen that the phase of the wavefunction rotates, and it is exactly reversed as we move to the opposite side of the contour\cite{Thomale}. Here we considered the kagome lattice with only the isotropic nearest neighbor hopping, but this feature remains even with distant hoppings.

\begin{figure}
	\includegraphics[width=7.5cm]{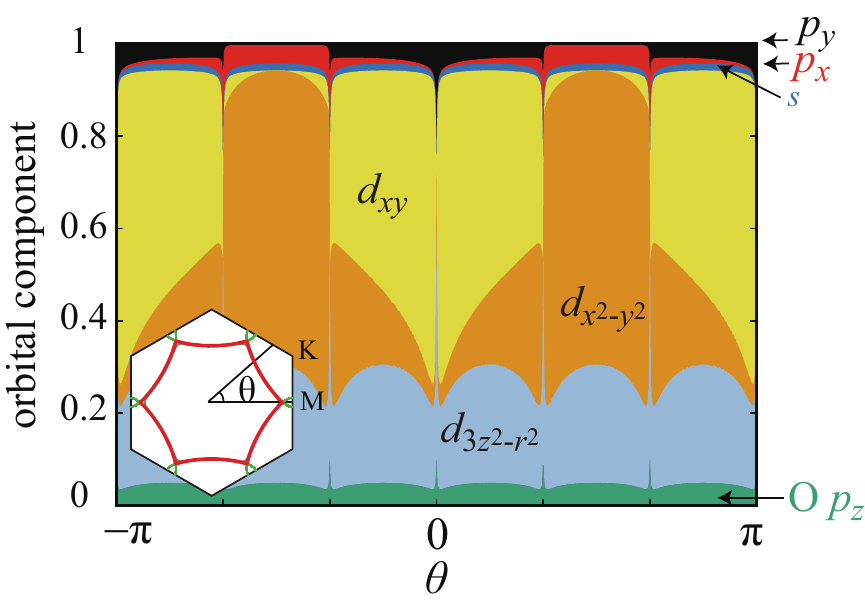}
	\caption{Weight of each orbital along the Fermi surface of PtCoO$_2$. Definition of the angle $\theta$ is given in the inset.}
	\label{fig9-2}
\end{figure}

We can say that this topological feature of the kagome-like electronic structure manifests itself as ``orbital-momentum locking'' in PtCoO$_2$ in the following sense. As mentioned above, the Fermi surface of PtCoO$_2$ mainly consists of the above mentioned Pt $d$ orbitals with a small amount of Pt $p$ and $s$ orbital mixture. In Fig. \ref{fig9-2}, we show how the orbital weight varies along the Fermi surface. This variance of the orbital character results in a ``rotation'' of the total wavefunction along the Fermi surface as shown in Fig. \ref{fig0}. Namely, a wave function, which looks somewhat like a well-known $d_{3z^2-r^2}$ orbital laid down in the $x-y$ plane, rotates along the Fermi surface. (To be more strict, the wavefunction has a different shape at the corners of the Fermi surface.) Therefore, at each point on the Fermi surface, the orbital character is different. 

We note that orbital-momentum locking in itself is not an extraordinary feature; it can in general be realized in multiorbital systems. For example, in a two orbital system originating from $p_x$ and $p_y$ orbitals (or $d_{xy}$ and $d_{x^2-y^2}$ orbitals) on a triangular lattice, the two orbitals give rise to two Fermi surfaces, and the mixture of the two orbitals results in a rotation of the $p$ orbital (or $d$ orbital) along each Fermi surface. Specific features of PtCoO$_2$ are that the Fermi surface essentially consists of only one band despite the strong multiorbital nature, and also that a very large Fermi velocity is realized due to the "hidden" mixture of $s + p_x/p_y$ orbitals. The effect of the orbital-momentum locking on the transport properties will be discussed in the next section.

Finally, let us point out an interesting feature regarding the kagome-like electronic structure observed above. In the original kagome lattice, the bands are two-fold degenerate at the K point (Fig. \ref{fig8}(a)), but in all of the kagome-like band structures hidden in the materials considered here, there is a (near) three-fold degeneracy, as seen in Fig. \ref{fig9} and Fig. \ref{fig8}(d). One may think that this is related to some kind of symmetry peculiar to the triangular lattice structure. However, we have found that this degeneracy can be lifted by employing artificial lattice structure parameters (e.g., varying the internal coordinate values of the oxygen atoms) without changing the symmetry of the triangular lattice structure. Also, this three-fold degeneracy is absent in the kagome-like electronic structure found in Na$_x$CoO$_2$ \cite{Koshibae}. At present, we are not certain about the origin of this interesting three-fold degeneracy.

\section{Impurity scattering}
\label{impurity}
In order to understand how the steep dispersion and the orbital-momentum locking are responsible for the extremely small resistivity observed in PtCoO$_2$, we evaluate the quasi-particle lifetime within the second order Born approximation \cite{Sheng}, which is justified for weak disorder. When the on-site disorder potential $U_0$ is uniformly distributed within the interval $(-U_0/2, U_0/2)$, the self-energy $\hat{\Sigma}$ due to disorder is defined through
\begin{equation}
	(E_F - \hat{H}_0 - \hat{\Sigma})^{-1} = \langle (E_F - \hat{H})^{-1} \rangle,  \label{eq_dis}
\end{equation}
where $\hat{H}_0$ is the Hamiltonian without disorder (i.e., the tight-binding Hamiltonian in this study), $E_F$ the Fermi energy, and $\hat{H}$ the Hamiltonian with the disorder potential, while $\langle\mathcal{O}\rangle$ denotes the disorder average of a operator $\mathcal{O}$.
$\hat{\Sigma}$ is self-consistently determined as follows,
\begin{equation}
	\hat{\Sigma} = \frac{1}{12}U^{2}_{0} \sum_{\bf k} \left[ E_F + i0^{+} - \hat{H}_0({\bf k}) - \hat{\Sigma} \right]^{-1}.
	\label{eq1}
\end{equation}
In the small $U_0$ regime, where the self energy is small, we can approximate Eq.~(\ref{eq1}) as 
\begin{equation}
	\hat{\Sigma} \simeq \frac{U^{2}_{0}}{12} \sum_{\bf k}\sum_{n} \frac{|n{\bf k}\rangle \langle n {\bf k}|}{E_F + i0^{+} - \varepsilon_{n}({\bf k}) - \langle n {\bf k} |\hat{\Sigma}| n {\bf k} \rangle },
	\label{eq2}
\end{equation}
where $|n{\bf k}\rangle$ is the wave function of $\hat{H}_0$ at wave vector ${\bf k}$ and the $n$-th band. This equation is solved self-consistently to obtain the self energy.
We then calculate its imaginary part
\begin{equation}
\tau_\theta^{-1}=\langle \theta |-{\rm Im} \hat{\Sigma}| \theta \rangle,
\end{equation} 
where $|\theta \rangle$ is the wave function $| n {\bf k} \rangle$ on the Fermi surface at the angle $\theta = {\rm tan}^{-1}(k_y/k_x)$.
This quantity corresponds to the scattering rate by impurities \cite{phonon}. 
We average $\langle \theta |-{\rm Im} \hat{\Sigma}| \theta \rangle$ over the Fermi surface and plot it against the disorder parameter $g = U^{2}_{0}/12$ (Fig. \ref{fig11}(a)).
 
\begin{figure}
	\includegraphics[width=5cm]{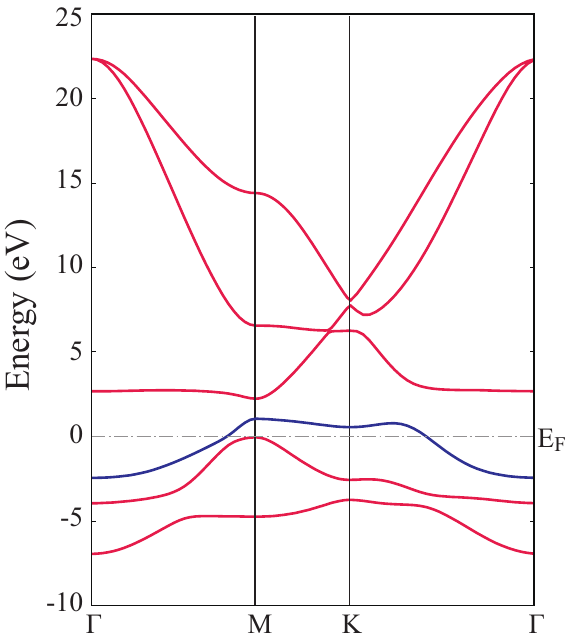}
	\caption{The band structure of the 6 orbital model with the spin-orbit coupling included. The blue line represents the dispersion of the single orbital model, which is obtained by extracting the band intersecting the Fermi level (see text).}
	\label{fig10}
\end{figure}

\begin{figure}
	\includegraphics[width=7cm]{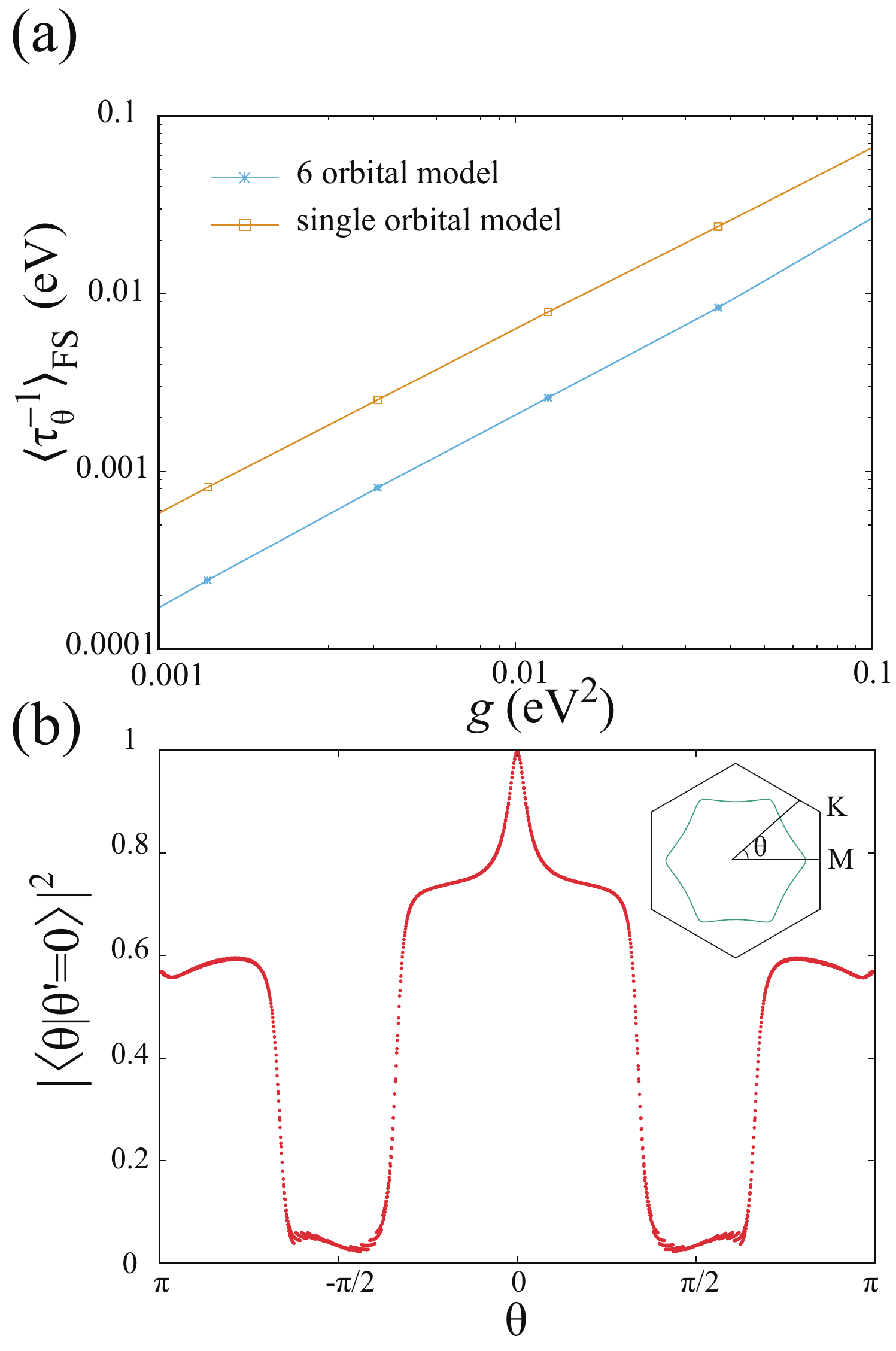}
	\caption{(a) The absolute value of the imaginary part of the self energy for the 6 orbital and the single orbital models. (b) The inner product of the wave functions on the Fermi surface plotted as a function of the angle $\theta$ measured from the $\Gamma$-M direction taken as $\theta'=0$.}
	\label{fig11}
\end{figure}

For the calculation of the self energy, we construct a 6 orbital model on a triangular lattice consisting of Pt $s$, $p_x$, $p_y$, $d_{3z^2-r^2}$ and $d_{xy}$ orbitals.
This 6 orbital model does not explicitly comprise the O $p_z$ orbitals (whose weight lies well below the Fermi level, see Fig. \ref{fig4}), but it accurately reproduces the band dispersion of the 8 orbital model near the Fermi level. It also appropriately considers the orbital components on the Fermi surface. Hence this model can be considered as a minimal model to investigate the impurity scattering. Here we further consider the spin-orbit coupling for the $d$-orbitals with a coupling constant of $\lambda = 0.75$ eV in order to get rid of the small pocket-like Fermi surface around the M point, so as to reproduce the experimental observations \cite{Kushwaha,KitamuraOka}. To be strict, we have found that the Pt $d_{xz}$, $d_{yz}$, $p_z$ orbitals mix significantly with the present orbtials when the spin-orbit coupling is turned on, and in this sense it is more accurate to use a 9 orbital model. However, we have checked this mixing of the additional three orbitals does not strongly affect the states near the Fermi level and hence the present calculation results. Therefore, here we concentrate on the 6 orbital model.
The Hamiltonian $\hat{H_0}$ in Eq. (\ref{eq_dis}) for the 6 orbital model is described as follows,
\begin{eqnarray}
    \hat{H}_0 &=& \hat{H}_{TB} + \hat{H}_{SO},\\
    \hat{H}_{SO} &=& \sigma_z \otimes \left(
    \begin{array}{cccccc}
      0 & -i\lambda & 0 & 0 & 0 \\
      i\lambda & 0 & 0 & 0 & 0 \\
      0 & 0 & 0 & 0 & 0 \\
      0 & 0 & 0 & 0 & 0 \\
      0 & 0 & 0 & 0 & 0 \\
      0 & 0 & 0 & 0 & 0
    \end{array}
  \right),
\end{eqnarray}
where $\hat{H}_{TB}$ contains the hoppings and the on-site energies of the 6 Wannier orbitals (orbital 1:$d_{xy}$, 2:$d_{x^2-y^2}$, 3:$d_{3z^2-r^2}$, 4:$p_x$, 5:$p_y$, 6:$p_z$), $\hat{H}_{SO}$ is the spin-orbit coupling term, and $\sigma_z$ is the $z$-component of the Pauli matrices.
Note that we consider only spin preserving scattering here.
We show in Fig. \ref{fig10} the band structure of this 6 orbital model with the spin-orbit coupling included, and in Fig. \ref{fig11} the imaginary part of the self energy calculated by using this model is plotted as a function of $g$.


In order to gain intuitive understanding, we now derive an approximate expression for the self energy. With the power series expansion for Eq. (\ref{eq1}), the imaginary part of the self energy can be described as follows, 
\begin{eqnarray}
	{\rm Im} \hat{\Sigma}  \sim -g\pi \sum_{{\bf k},n} \delta\left(E_F - \varepsilon_n ({\bf k}) \right)|n {\bf k}\rangle \langle n {\bf k}|.
	\label{eq3}
\end{eqnarray}
Because PtCoO$_2$ exhibits a single-band, strongly two-dimensional Fermi surface, we can describe Eq. (\ref{eq3}) with $|\theta \rangle$ and the distance from the $\Gamma$ point $k_F(\theta) = (k_x^2 + k_y^2)^{1/2}$,
\begin{eqnarray}
	{\rm Im} \hat{\Sigma} &=& -\frac{g S_0}{4\pi} 
	              \int_{0}^{2\pi}d\theta \int_{0}^{\infty} dk
		       \frac{k \delta\left(k - k_F(\theta)\right)}{v_{F}(\theta)} |\theta \rangle \langle \theta|,
\end{eqnarray}
where $S_0$ is the area of the unit cell within the $xy$ plane and  $v_{F}$ is the Fermi velocity at angle $\theta$. The self energy for angle $\theta$ is thus given by
\begin{eqnarray}
	{\rm Im} \langle \theta |\hat{\Sigma}| \theta \rangle \simeq
	   -\frac{gS_0}{4\pi} \int^{2\pi}_{0} d\theta' \frac{k_F(\theta')}{v_F(\theta')} |\langle \theta | \theta' \rangle|^2.
	   \label{eq4}
\end{eqnarray}
The overlap factor $|\langle \theta | \theta' \rangle|^2$ in Eq. (\ref{eq4}) is reduced from unity when states at $\theta$ and $\theta'$ have different orbital characters. Hence, Eq. (\ref{eq2}) shows that both the large group velocity near the Fermi level and the orbital-momentum locking reduce the imaginary part of the self energy, and thus lead to large conductivity.  

In order to highlight how orbital-momentum locking helps increase the conductivity, we also calculate, using Eq. (\ref{eq2}), the self energy for a single orbital model. In order to exclude the effect of the Fermi velocity, we consider a single orbital model that  has exactly the same energy dispersion as that of the band that intersects the Fermi level in the 6 orbital model, i.e., the band with the third-lowest energy (blue line in Fig. \ref{fig10}).  As shown in Fig. \ref{fig11}(a), the self energy of the single orbital model is much larger than that of the 6 orbital model. This result can be understood by looking into the inner product of the wave function on the Fermi surface for this model. As seen in Fig. \ref{fig11}(b), the scattering on the Fermi surface is strongly reduced, and about 70\% reduction of the self energy (the average of $|\langle \theta | \theta' \rangle|^2$ is about 0.3) is attained by the orbital-momentum locking.  We may hence conclude that the orbital-momentum locking  on the Fermi surface with 6 orbitals involved, in addition to the large Fermi velocity originating from the kagome lattice of $s+p_x/p_y$ orbitals, is the origin of the extremely small resistivity observed in PtCoO$_2$. 

The reduction mechanism for impurity scattering is reminiscent of that in  graphene. Namely, in graphene, the backward scattering of electrons by impurities is prohibited due to the pseudospin-momentum locking \cite{Ando}, where the pseudospin originates from the AB sublattices of the honeycomb lattice. In the present case, the multiorbital character plays the role of the pseudospins in graphene. From a topological viewpoint, in graphene the pseudospin-momentum locking is directly linked to the Berry phase of $\pi$ around the K point in graphene, whereas in the present case, the Berry phase of $2\pi$ around the two-fold degenerate band structure at the $\Gamma$-point, a feature of the kagome lattice, plays a similar role (see Fig. \ref{fig0}). The difference lies in that in the present case, the backward scattering$(\theta=\pi)$ is not strongly suppressed, but the scatterings around $\theta=\pi/2$, $3\pi/2$ are suppressed, reflecting the Berry phase of $2\pi$ instead of $\pi$. 

\section{Conclusion}
To conclude, the electronic structure of PtCoO$_2$ near the Fermi level is constructed from a mixture of Pt $s+p_x/p_y$ and Pt $d_{3z^2-r^2}+d_{xy}/d_{x^2-y^2}$ bands, both forming a hidden kagome-like electronic structure. The steep dispersion itself can give rise to a large mobility. In addition, these kagome-like features result in a mixture of six orbital characters on the Fermi surface, and the orbital-momentum locking, in addition to the steep dispersion itself, reduces the rate of the electron scattering by impurities. In fact, the experimentally observed Fermi velocity of $0.89\times 10^6$ m/s is comparable to, but not as large as those of very good metals such as copper or silver. Hence, the orbital-momentum locking is likely to be playing an important role in the realization of the extremely large conductivity. In total, we have concluded that the combined hidden kagome-like electronic structures, peculiar to the delaffosite compound, is the origin of the peculiar transport properties observed experimentally.

\section{ACKNOWLEDGMENTS}
We acknowledge F. Mazzola for useful discussions. 
We also acknowledge G. Fiete for fruitful discussions and a critical reading of the manuscript. We thank  H. Alloul and  I. Mukhamedshin for pointing out the relevance of refs.\cite{Alloul1,Alloul2,Lysogorskiy,AlloulRes} and also  R. Thomale for pointing out the relevance of ref.\cite{Thomale}. 
KK acknowledges fruitfull discussions with M. Koshino.
This work was supported by JSPS KAKENHI (Grants No. JP17K14108) and JST CREST (Grant No. JPMJCR16Q6), Japan, 
the European Research Council (Grant No. ERC-714193-QUESTDO), and The Royal Society.


\begin{thebibliography}{99}
\bibitem{phonon} One may also expect reduction of electron scattering by phonons, although we have not performed actual calculations on this problem.
\bibitem{commentkagome} We will use the term ``kagome-like electronic structure'' in this sense, although one may usually think of a band structure with one of the bands being perfectly flat.
\end{thebibliography}
\end{document}